\documentclass[12pt, preprint]{aastex}
\usepackage{graphicx}
\begin{document}


\title{Measuring the Spin of GRS 1915+105 \\with Relativistic Disk Reflection}

%% from AASTeX v4.0. You can use \email to mark an email address2

\author{J.L.Blum\altaffilmark{1}, J.M. Miller\altaffilmark{1}, A.C. Fabian\altaffilmark{2}, M.C. Miller\altaffilmark{3}, J. Homan\altaffilmark{4}, M. van der  Klis\altaffilmark{5}, \\
E. M. Cackett\altaffilmark{1,6}, R.C. Reis\altaffilmark{2}}

\altaffiltext{1}{Department of Astronomy, University of Michigan, 500 Church Street, Ann Arbor, MI 48109}
\altaffiltext{2}{Cambridge University, Institute of Astronomy, Madingley Road, Cambridge CB3 0HA, UK}
\altaffiltext{3}{Department of Astronomy, University of Maryland, College Park, MD 20742}
\altaffiltext{4}{Kavli Institute for Astrophysics and Space Research, Massachusetts Institute of Technology, 77 Massachusetts Avenue, Cambridge, MA 02139}
\altaffiltext{5}{University of Amsterdam, Astronomical Inst Anton Pannekoek, Kruislann 403, NL 1098 SJ Amsterdam, Netherlands}
\altaffiltext{6}{Chandra Fellow}

\begin{abstract}
GRS 1915+105 harbors one of the most massive known stellar black holes
in the Galaxy.  In May 2007, we observed GRS 1915+105 for $\sim$ 117
ksec in the low/hard state using \emph{Suzaku}.  We collected and
analyzed the data with the HXD/PIN and XIS cameras spanning the energy
range from 2.3$-$55 keV.  Fits to the spectra with simple models
reveal strong disk reflection through an Fe K emission line and a
Compton back-scattering hump.  We report constraints on the spin
parameter of the black hole in GRS 1915$+$105 using relativistic disk
reflection models.  The model for the soft X-ray spectrum (i.e.$<$ 10 keV)
suggests $\hat{a} = 0.56^{+0.02}_{-0.02}$ and excludes zero spin at
the 4$\sigma$ level of confidence.  The model for the full
broadband spectrum suggests that the spin may be higher, $\hat{a} =
0.98^{+0.01}_{-0.01}$ ($1\sigma$ confidence), and again excludes zero
spin at the $2\sigma$ level of confidence.  We discuss these results
in the context of other spin constraints and inner disk studies in GRS
1915$+$105.
\end{abstract}

%\keywords{}

\section{Introduction}

GRS 1915+105 was discovered by the WATCH instrument on board the {\it
Granat} satellite on 1992 August 15 (Castro-Tirado et al. 1992) .  It
has since been classified as a {\it microquasar}: a Galactic jet
source with properties similar to quasars, but on a stellar scale.
The mass of the central compact object is estimated to be 14$\pm$4
$M_{\sun}$ (Greiner et al. 2001).  The authors reported
results on the optical counterpart that implied a K-M III donating
star of $\sim$ 1.2 $M_{\sun}$, making GRS 1915+105 a low-mass X-ray
binary.  GRS 1915+105 is well-known for its extreme variability across
all bands of the electromagnetic spectrum (Belloni et al. 2000).

In 1998, BeppoSAX observations of GRS 1915+105 showed the clear
presence of a broad emission line at 6.4 keV (Martocchia et al. 2002),
i.e. the Fe K$\alpha$ line.  The Fe K$\alpha$ emission line is a
doublet consisting of the K$\alpha$1 and K$\alpha$2 lines at 6.404 keV
and 6.391 keV.  Such emission lines are likely produced by irradiation
of the accretion disk by an external source of hard X-rays (an
external X-ray source is required because accretion disks do not
efficiently self-irradiate).  It is generally thought that hard X-rays
originate in a hot, diffuse corona that inverse-Compton scatters soft
X-ray photons emitted by the accretion disk.  Irradiation of the dense
disk material by the hard X-rays then gives rise to a characteristic
``reflection" spectrum that results from Compton scattering and
photoelectric absorption (George \& Fabian 1991).  Photoelectric
absorption within the disk leads to the formation of the fluorescent
iron emission line at about 6.4 keV.  In stellar-mass black holes,
highly ionized iron lines are typically observed (Miller 2007).
Because of its large cosmic abundance and high fluorescent yield, the
Fe K$\alpha$ line is the most prominent line in the X-ray reflection
spectrum.

Disk reflection is an important diagnostic of black hole spacetime
geometry and general relativistic (GR) effects (e.g. Fabian 1989; Laor 1991; Brenneman \&
Reynolds 2006).  It is believed that the iron line originates in the
innermost parts of the accretion disk and thus is highly distorted by
Newtonian, special relativistic beaming, and GR
effects (e.g. Fabian, et al. 2000).  This skewing explains the broad,
skewed iron line profile that is observed in some stellar-mass X-ray
binaries, such as Cygnus X-1 and GRS 1915+105 (see Miller 2007).

%------------------------------------------------------------------------------------------------------------------------

In our modeling of the X-ray spectrum, we make the assumption
that the accretion disk extends down to the innermost stable circular
orbit (ISCO) in the low/hard state.  This may seem contrary to some models, in which the low/hard state is characterized by an inner disk boundary that is outside the ISCO.  However, several pieces of evidence suggest that our assumption is likely to be accurate for GRS 1915+105.  As follows, we observed GRS 1915+105 in the the low/hard state, which is characterized by an energy spectrum that is dominated by a power-law-like component.  While there is a typical photon index of $\sim$ 1.6 and a high-energy cutoff at $\sim$ 100 keV (Fender \& Belloni 2004), there is also a very weak soft X-ray component (most likely associated with a thermal disk) (Miller et al. 2006a).  According to Nowak (1995), the total X-ray luminosity for systems in the low/hard state where distance and mass estimates are known is generally below 10\% of the Eddington luminosity.   Esin et al. (1997), gave an archetypal model of how accretion flows change with state.  In that model, the inner disk in the low/hard state is radially truncated at low mass accretion rates (corresponding to $\mathrm{L_{X}/L_{EDD}}$ $\leq$ 0.008) and an advection-dominated accretion flow is present.  However, the luminosity observed in GRS 1915+105 (L $\sim$ 0.30 $\mathrm{L_{EDD}}$) is higher than what is theoretically calculated for a truncated disk in the low/hard state (Esin et al. 2007).  This higher luminosity, and therefore higher mass accretion rate, is indicative of inner disk extending closer to the black hole (e.g. Esin et al. 2007: Fig. 1) and supports the notion that radially recessed disks in the low hard state may not always be the case.  Recent analyses of GX 339-4, Cygnus X-1, SWIFT J1753.5-0127, and XTE J1817-330 have also revealed cool accretion disks that appear to remain close to the black hole at low accretion rates (Miller et al. 2006a, b, Rykoff et al. 2007, Tomsick et al. 2008).  

%----------------------------------------------------------------------------------------------------------------------

The radius of the ISCO is determined by the spin of the
black hole: $r_{\mathrm{ISCO}}$ = 6.0 $\mathrm{r_{g}}$ for
\emph{$\hat{a}$}=0, and $r_{\mathrm{ISCO}}$ = 1.25 $\mathrm{r_{g}}$
for an astrophysical maximum of \emph{$\hat{a}$} = 0.998 (where
$\mathrm{r_{g}}$ = GM/$\mathrm{c^{2}}$ and \emph{$\hat{a}$} =
cJ/G$\mathrm{M^{2}}$, see Thorne 1974, and Bardeen, Press, \&
Teukolsky 1972).  The location of the ISCO determines the ``strength''
of the gravitational effects on the Fe K emission line (and broadband
reflection spectrum), thereby influencing how broad and skewed the
line profile becomes.  Therefore, determining the ISCO through line
fitting serves as an indirect measure of the black hole spin
parameter.

Spectral fits to thermal emission from the accretion disk in black
hole binaries provides an independent way to constrain black hole
spin.  In order to turn the flux observed from the disk continuum into
a radius (and therefore a spin parameter), the source distance,
inclination, and mass must be known.  In addition, factors like the
line of sight absorption, hardening due to scattering in the disk
atmosphere, and the form of the hard spectral component can be
important.  In order to obtain reliable spin estimates using the continuum method, it is
crucial for the source to have a strong thermal component
(e.g. McClintock \& Remillard 2009).  For the case of GRS 1915+105,
this thermal component is most easily observed while in the high/soft
state.  However, prior spin results that have been reported for GRS
1915$+$105 in this state are not consistent.  McClintock et al. (2006)
and Middleton et al. (2006) both used models for thermal continuum
emission from the accretion disk to calculate values of
\emph{$\hat{a}$} $>$0.98 and $0.72^{+0.01}_{-0.02}$, respectively.  In
this paper, we present an independent analysis using the relativistic
disk reflection spectrum to calculate the black hole spin in GRS
1915+105 in the low/hard state with the advanced cameras on
\emph{Suzaku}.

\section{Data Reduction}

\emph{Suzaku} observed GRS 1915+105 on 2007 May 7 starting at
14:40:22 (TT).  The observation duration was approximately 117 ks.
The XIS pointing position was used.  In order to prevent photon
pileup, the XIS cameras were operated using the 1/4 window mode using
a 1.0 s burst option. The XIS1 and HXD/PIN cameras were used in this
analysis.  Two other XIS units were turned off to preserve telemetry
and the fourth unit was run in a special timing mode (which has yet to
be calibrated).  XIS on-source times of approximately 51 ks were
achieved.  This resulted in a dead-time corrected net exposure of 25
ks for the XIS1 camera (in 3x3 editing mode).  A net exposure time of
57 ks was achieved using the HXD/PIN.

For the soft x-ray data, the 2.3$-$10 keV energy range was used in order to avoid
calibration problems near the Si K edge and because there are few
photons at low energy owing to high line of sight absorption
($n_{\mathrm{H}}$ $\sim$ 4.0 x $10^{22} \mathrm{cm^{-2}}$).  Using the
clean event file (3 x 3 mode data) with the latest calibration
databases at the time from the XIS data (CALDB 20070731: version 2.0.6.13), we extracted the source
light curve and spectrum with \texttt{xselect}.  An annulus
centered on the source with an inner radius of 78$\arcsec$ and an
outer radius of 208$\arcsec$ (75 and 200 in pixel units,
respectively) was used for the source extraction region
because the center of the image suffered pile-up.  An annulus centered
on the source was also used for the background extraction region with
an inner radius of 208$\arcsec$ and an outer radius of 271$\arcsec$
(200 and 260 in pixel units, respectively).  We manually corrected for
extracted region areas that did not land on the XIS chip.
The XIS redistribution matrix files (RMFs) and ancillary response files
(ARFs) were created using the tools \texttt{xisrmfgen} and \texttt{xissimarfgen}
available in the HEASOFT version 6.6.2 data reduction package.  The 3 x 3 mode event file was used
to specify good time intervals and the data was grouped
to a minimum of 25 counts per bin using the FTOOL
\texttt{grppha}.

For the HXD data, the latest calibration databases (CALDB
20070710) were used and the reduction process began with a clean event file 
from the PIN detector (energy range 10.0$-$70.0
keV), which was at XIS aimpoint.  In the PIN data reduction process, the PIN
spectrum was extracted and deadtime was corrected by using the pseudo-events files.
After the non-X-ray (NXB) background spectrum was extracted, the exposure time of the background spectrum was increased by a factor of
10 since the NXB event file was calculated with a count rate 10 times
higher than the real background count rate to reduce statistical
errors.  The cosmic X-ray background was simulated, modeled and added to the NXB spectrum in
XSPEC version 12.5.0 as instructed in the \emph{Suzaku} ABC guide.  

\section{Data Analysis and Results }

Using the X-ray spectral fitting software package (XSPEC v. 12.5.0,
Arnaud 1996), we initially fit the \emph{Suzaku} XIS spectrum of GRS
1915$+$105 with a simple absorbed power law model: \texttt{phabs*(powerlaw)}, with the \texttt{phabs} model component accounting for Galactic photoelectric absorption.  All parameters were allowed to vary.  The 4.0$-$7.0 keV
range was ignored when fitting the data to obtain an unbiased view of the
Fe K range.  The fit resulted in an equivalent hydrogen column, $n_{\mathrm{H}}$, of $4.51_{-0.06}^{+0.06}$ (in units of $10^{22} $$\mathrm{cm^{-2}}$) and a photon index, $\Gamma$, of $2.09^{+0.01}_{-0.01}$ ($\chi^{2}$/$\nu$ = 1402/1282).  That ignored energy range was then restored when forming the
data/model ratio as shown in Figure 1.  An asymmetric, skewed line is
revealed in the exact way predicted for relativistic disk lines.  In a
similar manner, Figure 2 shows the XIS and HXD spectra fit with a
simple model consisting of a broken power-law modified by interstellar
absorption: \texttt{phabs*(bknpow)}.  A constant was allowed to float between the two data sets.  The 4.0$-$7.0 keV and 15.0$-$45.0 keV
ranges were ignored when fitting in order to properly model the continuum.  Note that for the HXD spectra, the model was fit over the 12.0$-$15.0 keV and 45.0$-$55.0 keV energy ranges.  The fit resulted in $n_{\mathrm{H}}$ = $4.35^{+0.06}_{-0.07}$, $\Gamma_{1}$ = $2.03^{+0.02}_{-0.02}$, and $\Gamma_{2}$ = $2.59^{+0.02}_{-0.02}$ with an energy break at $7.7^{+0.2}_{-0.2}$ keV ($\chi^{2}$/$\nu$ = 1565/1312).   Again, the ignored energy ranges were restored when forming the data/model
ratio. Note that the residuals near 12 keV can be attributed to calibration problems near the edge of the detector.  

XSPEC provides a number of models for spectral line fitting.  Models
of particular interest to us not only allow for analysis of the
observed broad iron line in our data, but account for the special and
GR effects that physically go into creating such an emission profile around a black hole.  Since determining the black hole spin of
GRS 1915+105 through analysis of the iron line is our main priority,
models that allow spin to be a free parameter in our fits are useful.
Brenneman \& Reynolds (2006) have calculated models (\texttt{kerrdisk}
and \texttt{kerrconv}) that describe line emission from an accretion
disk and include black hole spin as a free parameter, thereby allowing
us to formally constrain the angular momentum of the black hole and
other physical parameters of the system.  The \texttt{kerrdisk} model
describes line emission from an accretion disk, while the \texttt{kerrconv}
model allows one to convolve a reflection spectrum with the smeared
line function.

Although general relativity permits the spin parameter
\emph{$\hat{a}$} to have any arbitrary value, black holes can in
principle have spin parameters -1 $\le$ \emph{$\hat{a}$} $\le$ 1.
However, for simplicity the \texttt{kerrdisk} model only considers a
black hole that has a prograde spin relative to the accretion disk
that spins up to the Thorne (1974) spin-equilibrium limit, i.e., 0
$\le$ \emph{$\hat{a}$} $\le$ 0.998.

In addition to the spin parameter, we can specify and/or constrain
nine other physical parameters in the \texttt{kerrdisk} model: (1)
rest frame energy of the line, (2) emissivity index within a specified
radius, (3) emissivity index at radii larger than a specified value,
(4) break radius separating the inner and outer portions of the disk
(in units of gravitational radii), (5) the disk inclination angle to
the line of sight, (6) the inner radius of the disk in units of the
radius of marginal stability (rms), (7) the outer radius under
consideration, (8) the cosmological redshift of the source (\emph{z}=0
for our source), and (9) the normalization (flux) of the line in
photons $\mathrm{cm^{-2}}$ $\mathrm{s^{-1}}$.  \texttt{kerrconv} has the same basic set of parameters with
the exception that it does not require an input line energy or a flux
normalization parameter since it uses a \texttt{kerrdisk} kernel to
smear the entire spectrum with relativistic effects.  Of these parameters, we let 1, 2, 3, 5, and 9 vary
freely.  We bounded the rest frame energy for the iron line between
6.4 and 6.97 keV and the disk inclination between $55\,^{\circ}$ and
$75\,^{\circ}$ based on the figure presented in Fender et al. (1999: Fig. 6) from
radio jet observations.  The emissivity indices were fixed to be equal
to each other meaning that we only consider one index, making the
break radius meaningless.  Parameters 4, 6 and 7 were frozen at 6.0
$\mathrm{r_{g}}$, 1.0 rms, and 400 rms, respectively using the
``standard" values of the \texttt{kerrdisk} and \texttt{kerrconv}
models.  By doing this, we assume that the ionization of the disk
material within the ISCO is too high to produce significant line
emission as has been shown by 3D MHD simulations (Reynolds \& Fabian
2008).

As shown in Fig.3, our model is defined as:
\texttt{phabs*(kerrdisk + kerrconv*pexriv)}.  A constant was allowed to float between the XIS and HXD data sets.  The model component
\texttt{pexriv} is an exponentially cut-off power-law spectrum
reflected from ionized material (Magdziarz \& Zdziarski 1995). The
free parameters are the power-law photon index $\Gamma$, the fold
energy or cutoff energy (in keV), and the scaling factor for
reflection.  Our results are given in Table 1 with the model being fit in the 2.3$-$10.0 keV and 12.0$-$55.0 keV energy ranges.  Note that the residual feature at the upper limit of the XIS detector is the result of noise.  
Errors were calculated using the
\emph{steppar} command in XSPEC, which affords control over how the
$\chi^{2}$ space is searched.  For all parameters, errors refer to the
68$\%$ confidence level ($\Delta$$\chi^{2}$= 1) unless otherwise
stated.  The reduced $\chi^{2}$ of our best-fit is 2345/2224.  We measure a spin value of $\hat{a} = 0.98_{-0.01}^{+0.01}$.
Figure 5 plots the dependence of $\chi^{2}$ on the black hole spin
parameter.  A spin of zero is excluded at the $2\sigma$ level of
confidence.

%-----------------------------------------------------------------------------------------------
The disk ionization parameter ($\xi$ = $\mathrm{L_X/nr^{2}}$, where n is the hydrogen number density) has a large effect on the resulting spectrum.  Ross et al. (1999) demonstrated  the effect on reflection spectra models for a range of ionization values ( 30$<$ $\xi$ $<$ $10^{5}$).  The model with the highest ionization parameter, $\xi$ = $10^{5}$ was the best reflector but had negligible iron spectral features due to the disk surface layer being fully ionized at great depth, with the iron line (specifically, Fe XXVI) not becoming dominant until $\tau$ $\approx$ 8.  However, when $\xi$ is reduced to 3 x $10^{3}$, less than half of the iron is fully ionized at the disk's surface and Fe XXV becomes dominant at $\tau$ $\approx$ 1.  Most importantly, at this value an iron emission line that is Compton-broadened becomes visible (Ross et al. 1999: Fig 2).  Similar to this ionization value is our value of 5000, which is the upper limit allowed by the \texttt{pexriv} model.   This ``high'' value was chosen so as to obtain the maximum amount of broadening due to Comptonization and further understand the dynamics of the system.  In our model, maximizing or minimizing the disk temperature ($10^{6}$ and $10^{4}$ Kelvin, respectively) changed $\chi^{2}$ by less than 5, resulting in similar reduced $\chi^{2}$ to the best-fit value ($\chi^{2}$/$\nu$ $\sim$ 1.05).  Effects to the parameter values (most notably the spin parameter) were negligible with the exception of the reflection scaling factor, which changed by $<$ 0.06.  Note that a 
thermal disk component is not required for our model and, since the disk
itself is likely to be relatively cool and the high Galactic
photoelectric absorption prevents its detection, we have set a low disk temperature (see, e.g., Miller et
al.\ 2006, Rykoff et al.\ 2007).

Given the complexity of the source and it's extreme variability, we had set the iron abundance equal to the solar value to provide a more simplistic approach to constraining the spin, our main parameter of interest.  Using the upper limit on the ionization value we tested the effects of altering the iron abundance in the \texttt{pexriv} model.  Ross \& Fabian (2005: Fig. 5) show that increasing the iron abundance relative to the solar value serves to lower the continuum while enhancing the Fe K$\alpha$ emission lines.  Additionally, Lee et al. (2002) measured an iron abundance, Fe/solar, $>$ 1.0 for GRS 1915+105, conveying that the overabundance may be due to dynamics of supernova in the microquasar's history.   However, Ueda et al. (2009) were more definitive in fitting a spectrum with better statistics using solar values for the abundances (including iron).  As such, after testing the effects of varying abundance, our \texttt{pexriv} model did not require an enhanced abundance (e.g. tripling the iron abundance increased $\chi^{2}$ by $\sim$ 100).   
%---------------------------------------------------------------------------------------------------------------------------------------------------

Our broadband model uses \texttt{pexriv} as its main
reflection component.  One difficulty with this model is that it does
not include the effects of Comptonization on photoelectric absorption
edges in the outer, most highly ionized layers of the disk.  Being statistically inferior in this regime, it may not be ideal to use this model for high values of $\xi$, which is the reason our value of 5000 erg cm/s is the upper limit.  In order to check how our spin constraints might depend on the
reflection model, we replaced \texttt{pexriv} with a newer, more self-consistent reflection model: \texttt{reflionx} (see Fig. 4).  The  \texttt{reflionx} model is a revised version of reflion (Ross \& Fabian 2005) that describes reflection from accretion disks in black hole systems where the blackbody emission is at too low an energy to affect the Fe K$\alpha$ emission.  It has been used previously to constrain the spin in other sources such as GX 339-4 (Reis et al. 2008).  The parameters of the model are the iron abundance (set to solar), photon index of the illuminating power law, ionization parameter, $\xi$, and the normalization.  In fitting \texttt{reflionx}, the HXD spectrum was neglected
because a cut-off power-law is required whereas \texttt{reflionx}
assumes that a simple power-law spectrum irradiates the disk.
Moreover, \texttt{reflionx} is a pure reflection model that requires a
simple power-law to represent the continuum.  Note that the Fe K emission line is self-consistently modeled within the \texttt{reflionx} model.  Also, in order to
obtain better constraints within the \texttt{reflionx} model, 
the inclination was bounded between at $55\,^{\circ}$ and $75\,^{\circ}$. 
The model was fit in the 2.3$-$10.0 keV energy range.  Letting $\hat{a}$ be a free parameter, the
\texttt{phabs*(powerlaw+kerrconv*reflionx)} model resulted in a spin
parameter of 0.56 with a $\chi^{2}$ value of approximately 2124 and a
reduced $\chi^{2}$ $\sim$ 1.01 for 2100 degrees of freedom (see Table 2).  Based on these results
from the \texttt{reflionx} model, we can state that $\hat{a} = 0.56^{+0.02}_{-0.02}$
is preferred and that a spin of zero is excluded at the $4\sigma$ level
of confidence (see Fig. 6).  However, it is important to note that the parameter values obtained with the \texttt{reflionx} model are poorly constrained, which requires the \texttt{pexriv} results to be viewed with extra caution (see Sect. 4).

Finally, we note the possible presence of an absorption line at
approximately 7.4~keV (see Figure 1).  It is not clear that the
feature is real.  When the broadband spectrum is fit
phenomenologically, a Gaussian model for the line suggests a
(single-trial) significance of $4\sigma$.  The feature is not
significant after fitting the relativistic emission line and disk
reflection continuum.  If the line is real, it could plausibly be
associated with Fe XXV or Fe XXVI, implying a wind with a blue-shift
as high as $0.1c$.  When a strong line is frozen as part of the
overall spectral model, even modest constraints on the spin parameter
of the black hole cannot be obtained.  However, a wind in a hard state
would be inconsistent with an apparent anti-correlation between winds
and jets in GRS 1915$+$105 (Miller et al.\ 2008a, Nielsen \& Lee
2009).
  
\section{Discussion}

In this paper, we have presented the first results from \emph{Suzaku}
observations of GRS 1915$+$105 in a low/hard state.  We have observed
the Fe K emission line in GRS 1915+105 in unprecedented detail due to
the spectral resolution and fast readout modes of the XIS cameras.
Our broadband spectral model suggests a spin of $\hat{a} =
0.98^{+0.01}_{-0.01}$, though a value of zero is only excluded at the
$2\sigma$ level of confidence.  A different model, fitted only to the soft X-ray spectrum, results in a spin of 
$\hat{a} = 0.56^{+0.02}_{-0.02}$ with a value of zero excluded at $4\sigma$ level of confidence.  This effort to measure the spin of the
black hole in GRS~1915$+$105 follows other recent efforts to measure
spin in stellar-mass black holes using relativistic iron lines (Miller
et al.\ 2008b, Reis et al.\ 2008/2009, Miller et al.\ 2009) and the thermal
disk continuum spectrum (e.g. Shafee et al.\ 2006).  

Owing to its extreme behavior and the possibility of a relation to
the spin of the black hole, it is particularly important to understand
the black hole in GRS 1915$+$105.  Middleton et al. (2006) and
McClintock et al. (2006) used similar thermal emission continuum
models based on different data selections from \emph{RXTE} to estimate
the spin of the black hole in GRS 1915$+$105.  Middleton et al. (2006)
determined the spin of GRS 1915+105 from a set of 16 \emph{RXTE}
observations (approximately spanning 1994 to 1996) that were 16 s
each.  The authors chose spectra where there was a slow
change between spectral states, specifically variability classes
$\beta$ and $\lambda$ (Belloni et al. 2000), which is why short
time-scale spectral binning was used.  They
required that Comptonization contribute less than 15 \% of the
bolometric flux (or rather that the disk contributed greater than 85
\%).  Also, the rms variability had to be $<$ 5 \%.  For the final
analysis, there were a total of 34 disk dominated spectra for 6
observations.  The XSPEC models used were for a multicolor disk
blackbody (\texttt{diskbb}), relativistic effects (\texttt{bhspec}),
and thermal Comptonization (\texttt{thcomp}) that result in an
intermediate spin value of \emph{$\hat{a}$} = $0.72^{+0.01}_{-0.02}$

For comparison, McClintock et al. (2006) used a much larger sample of
\emph{RXTE} data than Middleton et al. (2006) with a total observation
time of 89 ks for \emph{RXTE} and 13 ks for \emph{ASCA}.  The 22
observations selected had to meet 3 criteria: (1) weak QPOs, (2) rms
continuum power $<$ 0.075 rms, and a (3) disk flux $>$ 75 \% of the
total 2-20 keV emission.  The authors used a relativistic
disk model (\texttt{kerrbb2}: a combination of \texttt{kerrbb} and
\texttt{bhspec} that included a spectral hardening factor), which
would alternately be combined with a power-law, Comptonization
(\texttt{comptt}), and a cut-off power-law model (\texttt{cutoffpl}).
These last 3 models represented the nonthermal tail component of the
emission.  They argued that the spin
(\emph{$\hat{a}$} $>$ 0.98) and mass accretion rate parameters were
relatively unaffected by the model for the nonthermal tail component.

Both sets of authors have identified several reasons that might
explain the different spin values that they derived.  McClintock et
al. (2006) note the existence of uncertainties in disk structure at
high luminosities.  These uncertainties could possibly affect the
Middleton et al. (2006) results, which used solely high luminosity
observations (L $>$ 0.3 $\mathrm{L_{EDD}}$).  McClintock et al. (2006)
also suggest that Middleton et al. (2006) missed crucial low
luminosity observations because of small data samples -- the derived
intermediate spin value may be due to averaging over a wide range of
luminosities (including a super-Eddington luminosity of L $\sim$ 1.45
$\mathrm{L_{EDD}}$).  However, Middleton et al. (2006) suggest that
the McClintock et al. (2006) low luminosity data (L $<$ 0.3
$\mathrm{L_{EDD}}$) does not allow for a low temperature Comptonized
component (k$\mathrm{T_{e}}$ $<$ 3 keV).

In summary, McClintock et al.  (2006) finds a rapidly spinning black
hole in GRS 1915+105 based on \emph{disk continuum} fits from
\emph{RXTE} and \emph{ASCA} data.  Middleton et al.  2006) used
similar assumptions (i.e. d $\sim$ 12 kpc and inclination angle $\sim$
$66\,^{\circ}$ (Fender et al. 1999)) and models, but had a smaller
sample of \emph{RXTE} data of GRS 1915+105 and got an intermediate
value of $\hat{a} = 0.72^{+0.01}_{-0.02}$.  The behavior of GRS
1915$+$105 is particularly complex and the states identified in this
source are more numerous and nuanced than those identified in other
black hole binaries.  The differences in Middleton et al. (2006) can
be partly attributed to the state classifications (i.e. A, B, C) for
GRS 1915+105 as opposed to the black hole binary categories used by
McClintock et al. (2006).  The spin estimate reported by McClintock et
al. (2006) was obtained in a state that resembles the ``thermal state"
(i.e. high/soft state or ``thermal dominant" state).  Here, our
observation was obtained in the low/hard state (which is mostly likely
``state C'').

Previous work has suggested that for non-spinning black holes, the
amount of Fe K emission from within the ISCO is negligible: inside the
ISCO, the ionization fraction sufficiently increases close to the
black hole making it unlikely that there is significant Fe K emission
within the ISCO since the ionization fraction substantially increases
(Reynolds \& Begelman 1997).  Reynolds \& Fabian (2008) find that
systematic errors on the spin due to emission from within the ISCO
become smaller for more rapidly rotating black holes.  Additionally,
Shafee et al. (2008) determined that magnetic coupling is unimportant
across the ISCO for geometrically thin disks.  The best current
theoretical works suggests that systematic errors are unlikely to bias
our results for GRS 1915+105 (Reynolds \& Fabian 2008).
 
An $\mathrm{r^{-3}}$ disk emissivity relation is typically assumed for
line emission from a standard thin accretion disk with an isotropic
source of disk irradiation (see, e.g, Reynolds \& Nowak 2003).  Freezing the emissivity index at 3.0 in the \texttt{pexriv} and \texttt{reflionx} models resulted in worse fits with $\Delta$$\chi^{2}$ $\gtrsim$ 200 for 1 degree of freedom (F-value $>$180, $>$ 8.0$\sigma$ level of confidence).  Our current best-fit value for the emissivity index in GRS
1915+105 is of interest because it implies that the reflected emission
has a weaker dependence on the radius ($\mathrm{r^{-2.0}}$) than is
typical.  When the emissivity index is high, as in the case for many spinning active
galactic nuclei, there is significant iron line flux below 4 keV from material closest to the black hole and therefore 
high gravitational redshift.  This is not true when the emissivity index is low.  Thus, spin determination for GRS 1915+105 case
is being driven more by the detailed shape of the body of the iron line rather
than the overall line extent.  This could explain much of the difference between the \texttt{pexriv} and \texttt{reflionx} results.  
Moreover, a stable warp at the inner disk (e.g. due to Lense-Thirring
precession) is a possible explanation for our different emissivity
index value in that it would reflect more than expected at larger
radii.  A link between Fe K emission and the phase of low-frequency
quasi-periodic oscillations (QPOs) in the low/hard state of GRS
1915$+$105 may provide evidence for such a warp (Miller \& Homan
2005).  

Martocchia et al. (2006) made two \emph{XMM-Newton} observations of
GRS 1915+105 in 2004 during a long ``plateau" (lower flux) state.  The
data was analyzed with EPIC-pn in timing and burst modes,
respectively.  For the timing mode, they detected a broad excess at
the energy of the iron line and fit the data from 2.0$-$10.0 keV with
the disk reflection model \texttt{diskline} (Fabian et al. 1989).
Martocchia et al. (2006) found an inner radius of the disk in excess
of 300 $\mathrm{R_{g}}$ for an unabsorbed flux of 1.0$\pm$0.1 ($10^{-8}$
$\mathrm{erg/s/cm^{2}}$) in the 2.0$-$10.0 keV energy range.  Although such a disk would have the
advantage of providing a geometric explanation for the major X-ray
spectral transitions seen in Galactic black holes (Done \&
Gierli\'{n}ski 2006), the high (unabsorbed) flux ($\sim$ 3.3 x $10^{-8}$
$\mathrm{erg/s/cm^{2}}$) and hence high luminosity ($\sim$ 0.30
$\mathrm{L_{EDD}}$) we calculated in the 0.5$-$100 keV range for GRS
1915+105 make it unlikely that the disk would be recessed to such a
high value (e.g. Esin et al. 1997, Frank, King \& Raine 2002, also see Miller et
al.\ 2006a).  We note that the high flux of GRS 1915+105 may have
caused photon pile-up in the EPIC-pn timing mode, thereby obscuring
the breadth of the line and affecting the Martocchia et al. (2006)
results.

Based on VLA observations of relativistic ejections from GRS 1915+105,
$70\,^{\circ}\pm2$ is the typical value used for the inclination of
this source (Mirabel \& Rodr\'{i}guez 1994).  However, in our broadband
model the inclination angle pegged at $55\,^{\circ}$, which is
close to the lowest plausible value for this black hole binary (we set
the bounds between $55\,^{\circ}$ and $75\,^{\circ}$, Fender et
al. 1999: Fig. 6).  Allowing the inclination to vary over a greater range ($20\,^{\circ}$$-$ $80\,^{\circ}$) resulted in our broadband model finding the minimum $\chi^{2}$ at $40\,^{\circ}$, which we disregard on the basis that it is not consistent with previous radio jet observations (e.g. Mirabel \& Rodr\'{i}guez 1994, Fender et al. 1999).  Given our assumption that the spin axis of the black
hole is parallel to the jet axis, our best-fit spin and inclination results
follow from self-consistent assumptions.  

As mentioned previously, the resulting parameter values in the \texttt{reflionx} model are not well constrained.  To better demonstrate the dependence between parameters, particularly for the spin parameter, figures 7 and 8  show the 68, 90 and 99 \% contour plots for inclination versus spin for both the \texttt{reflionx} and \texttt{pexriv} models, respectively.   As shown for the \texttt{reflionx} model, there is a large degree of variation between the inclination and spin values whereas for the \texttt{pexriv} model, the spin and inclination values are more restricted.  Due to the poor constraints in the \texttt{reflionx} model, the \texttt{pexriv} model results must be regarded with more caution.

Among stellar-mass black holes, GRS 1915$+$105 displays uniquely rich
and complex phenomena, both in X-rays and at shorter wavelengths.  Its
X-ray spectrum is similarly complex, and whether using the disk
continuum or the reflection spectrum to constrain the spin, data
seletion and modeling nuances can be important.  Currently, our broadband model (\texttt{pexriv}) gives a spin parameter
that is in agreement with the value reported by McClintock et
al. (2006) (\emph{$\hat{a}$} $>$ 0.98), while the upper limit on our
soft X-ray spectrum model (\texttt{reflionx}) spin result is lower than
the Middleton et al. (2006) spin value.   In this effort to
constrain the spin of GRS 1915$+$105, we have considered only a single
observation in a single state, though disk reflection is sure to be
important in many states.  Future observations with {\it Suzaku} and
{\it XMM-Newton} in different states may allow the spin to be
determined more clearly using the disk reflection spectrum.  The
thermal disk continuum is not evident in our observation, but future
efforts that make joint use of the direct and reflected disk spectrum
to constrain the spin (e.g. Miller et al.\ 2009) may provide another
way forward.

\section*{Acknowledgements}
\noindent
We gratefully acknowledge the anonymous referee for helpful comments that improved this work.  We thank the Suzaku mission managers and staff for executing our TOO
observation.  We thank Koji Mukai for helpful conversations.  J.M.M.
acknowledges funding from NASA through the \emph{Suzaku} guest
investigator program.  M.C.M. acknowledges funding from NSF grant
AST0708424.  E.M.C.  gratefully acknowledges support provided by NASA
through the \emph{Chandra} Fellowship Program, grant number PF890052.
R.C.R. thanks STFC for grant support.  This work has made use of the tools and facilities available through
HEASARC, operated for NASA by GSFC.

%------------------------------------Table Start-----------------------------------------------------------

\begin{deluxetable}{l l l}
\tablecolumns{3}
\tablewidth{0pc}
\tablecaption{XSPEC model}
\tablehead{\colhead{Model component}                   &                                \colhead{Parameter}                           &             \colhead{Value}}
\startdata              
             \texttt{phabs}   				&      				nH ($10^{22}$) ($\mathrm{cm^{-2}}$)        &                $4.15^{+0.06}_{-0.06}$      \\
             \texttt{kerrdisk}				&			         Fe line energy  (keV) 		                   &                 $6.40^{+0.06}$        \\
                							&                                   rb (rms)							&	          	(6.0)\\
					               		&			        Emissivity Index                                                                  &                 $1.8^{+0.1}_{-0.1} $       \\
                 							&				a ($\mathrm{cJ/GM^{2}}$)                             &                 $0.98^{+0.01}_{-0.01} $     \\
              							&			         Inclination (deg)                 				 &                 $55.0^{+2.0}   $     \\
							         &                                  $\mathrm{R_{i}}$                                              &                   (1.0)\\							      
							          &                                  $\mathrm{R_{o}}$                                            &                     (400)\\
							         &				Redshift                                                             &                     (0)\\							        
							          &			         Norm.  ($\mathrm{photons/cm^{2}/s}$)         &                 $0.010^{+.002}_{-.002}  $   \\
             \texttt{pexriv}					&			         Photon Index                       				  &                 $1.96^{+0.03}_{-0.03}   $   \\
               							&			         Fold E (keV)                      				  &                  $38^{+1.0}_{-1.0}   $  \\
                							&				Rel. Reflection                 				   &                  $0.38^{+0.02}_{-0.02}   $ \\
               							&                                   Fe/solar							&			(1.0)\\
							         &                                   Cosine of Inc. Angle                                      &                           (0.45)\\
							         &				$\mathrm{T_{disk}}$ (K)                                     &                         (300,000)\\
							         &                                   Disk Ionization (erg cm/s)                                 &                   (5000)\\
							         &				Norm. ($\mathrm{photons/keV/cm^{2}/s}$)   &                  $3.8^{+0.10}_{-0.10}  $ \\
                                                &																		&										\\
 \texttt{$\chi^{2}/\nu$}				&		                                                                             &                       2345/2224        \\				          
\enddata
\tablecomments{These are the best$-$fit parameters found using the model \texttt{phabs}*(\texttt{kerrdisk}+\texttt{kerrconv}*\texttt{pexriv}) with the XIS and HXD data.  All of the \texttt{kerrdisk} and \texttt{kerrconv} parameters were tied and constant was allowed to float between the XIS and HXD data sets.   The errors quoted in Table 1 are 1$\sigma$ errors unless otherwise noted.  The values in parentheses were fixed during the fitting. The calculated luminosity (0.5$-$100 keV) ($\mathrm{10^{38} erg/s}$) for a distance d $\sim$ 12 kpc (Fender \& Belloni 2004) is $5.7^{+0.10}_{-0.10}$. }
\end{deluxetable}

%----------------------------------Table End------------------------------------------------------------

%------------------------------------Table Start-----------------------------------------------------------

\begin{deluxetable}{l l l}
\tablecolumns{3}
\tablewidth{0pc}
\tablecaption{XSPEC reflionx model}
\tablehead{\colhead{Model component}                   &                                \colhead{Parameter}                           &             \colhead{Value}}
\startdata              
             \texttt{phabs}   				&      				nH ($10^{22}$) ($\mathrm{cm^{-2}}$)        &                $5.64^{+0.03}_{-0.03}$      \\
             \texttt{powerlaw}				&			         Gamma 		                                               &                 $2.44^{+0.01}_{-0.01}$        \\
                							&                                   Normalization			           		&	          	$6.78^{+0.03}_{-0.03}$\\
             \texttt{kerrconv}					&			         Index                       				  &                 $2.1^{+0.1}_{-0.1}$   \\
               							&			         rb (rms)                      				  &                  (6.0)  \\
                							&				a ($\mathrm{cJ/GM^{2}}$)                				   &                  $0.56^{+0.02}_{-0.02}$ \\
               							&                                   Inclination (deg)							&	$69.3^{+2.0}_{-2.0}$ 		\\
							         &				$\mathrm{R_{i}}$						&                           (1.0)\\
							         &                                    $\mathrm{R_{o}}$                                 &                         (400)\\
	\texttt{reflionx}			                 &                                   Fe/solar                                 &                   (1.0)\\
							         &				Gamma				   &                  $2.44^{+0.01}_{-0.01}$ \\
                                             			   &					Xi													&	$100^{+3.0}_{-3.0}$									\\
			   				&				Redshift                                  &               (0)      \\
							&                          Normalization ($10^{-3}$)     &          $2.5^{+0.3}_{-0.3}$            \\	 \texttt{$\chi^{2}/\nu$}			&		                                                                             &                       2124/2100        \\				          
\enddata
\tablecomments{These are the best$-$fit parameters found using  the \texttt{phabs*(powerlaw+kerrconv*reflionx)} model with only XIS data.  The errors quoted are 1$\sigma$ errors unless otherwise noted.  The values denoted in parentheses are fixed parameters. }
\end{deluxetable}

%----------------------------------Table End------------------------------------------------------------

%------------------------------------Figure Start---------------------------------------------------------
\begin{figure}[t]
\begin{center}
\includegraphics[scale=0.40]{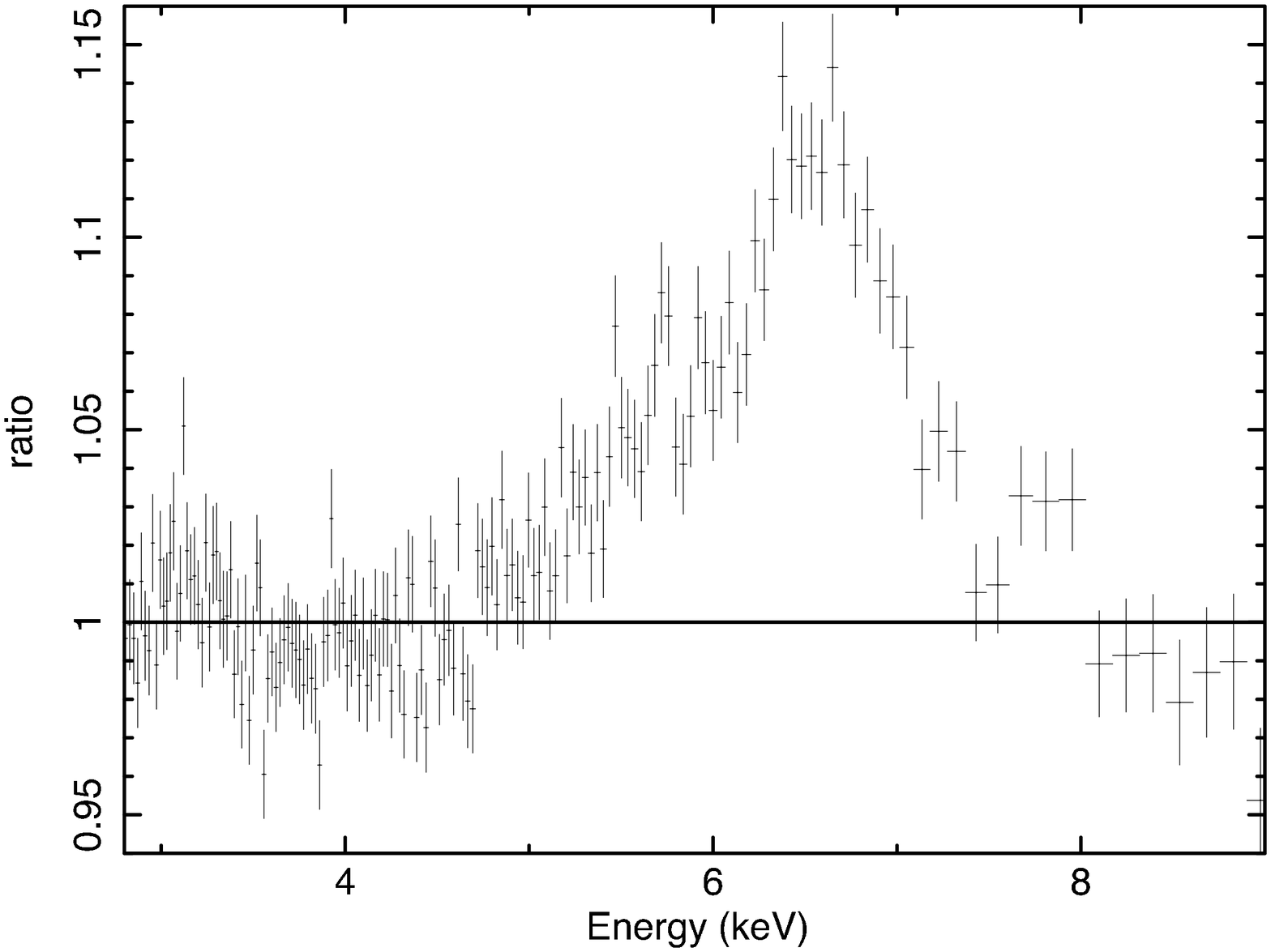}
\vspace{-1.0cm}
\end{center}
\caption{\footnotesize
{ Data/model ratio obtained when the XIS 1 \emph{Suzaku} spectrum of GRS 1915+105 was fit with a simple power-law model: \texttt{phabs*(powerlaw)}, that included photoelectric absorption.  The 4.0$-$7.0 keV region was ignored when fitting the model.  An asymmetric skewed line profile is evident in the way predicted for relativistic lines.}}
\label{fig1}
\end{figure}

%-----------------------------------Figure End---------------------------------------------------------

%------------------------------------Figure Start---------------------------------------------------------
\begin{figure}[t]
\begin{center}
\includegraphics[scale=0.40]{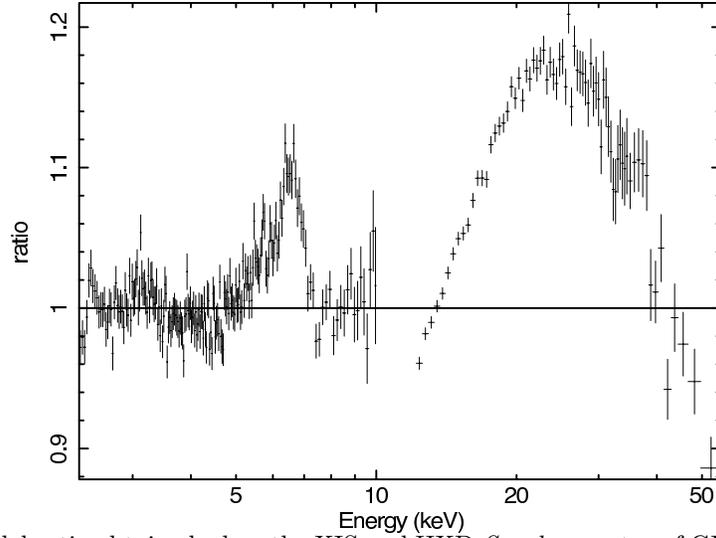}
\vspace{-1.0cm}
\end{center}
\caption{\footnotesize
{ Data/model ratio obtained when the XIS and HXD \emph{Suzaku} spectra of GRS 1915+105 were fitted with a broken power law model: \texttt{phabs*(bknpow)}, that included photoelectric absorption.  For the HXD spectra, the model was fit over the 12.0$-$15.0 keV and 45.0$-$55.0 keV energy ranges.  The 4.0$-$7.0 keV and 15.0$-$45.0 keV regions were ignored when fitting the model.  The residuals near 12 keV can be attributed to calibration problems near the edge of the detector.  The curvature at high energy is a clear signature of disk reflection.}}
\label{fig2}
\end{figure}

%-----------------------------------Figure End---------------------------------------------------------

%------------------------------------Figure Start---------------------------------------------------------
\begin{figure}[t]
\begin{center}
\includegraphics[scale=0.40]{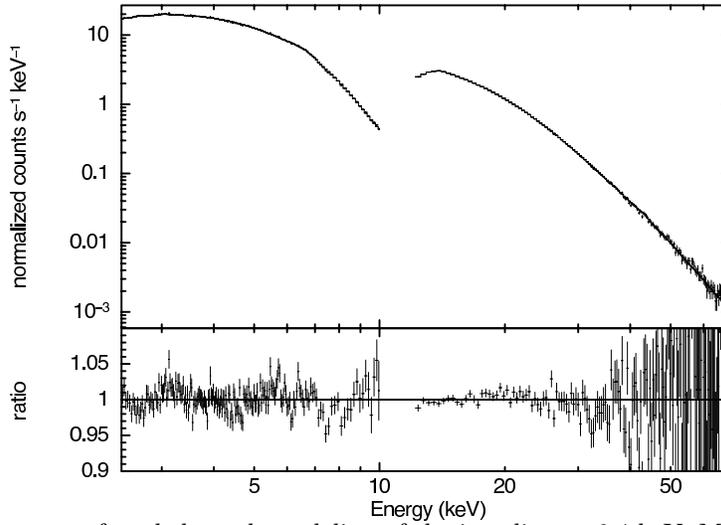}
\vspace{-1.0cm}
\end{center}
\caption{\footnotesize
{Best-fit spectrum found through modeling of the iron line at 6.4 keV.  Model: \texttt{phabs*(kerrdisk + kerrconv*pexriv)}. }}
\label{fig3}
\end{figure}

%-----------------------------------Figure End---------------------------------------------------------

%------------------------------------Figure Start---------------------------------------------------------

\begin{figure}[t]
\begin{center}
\includegraphics[scale=0.40]{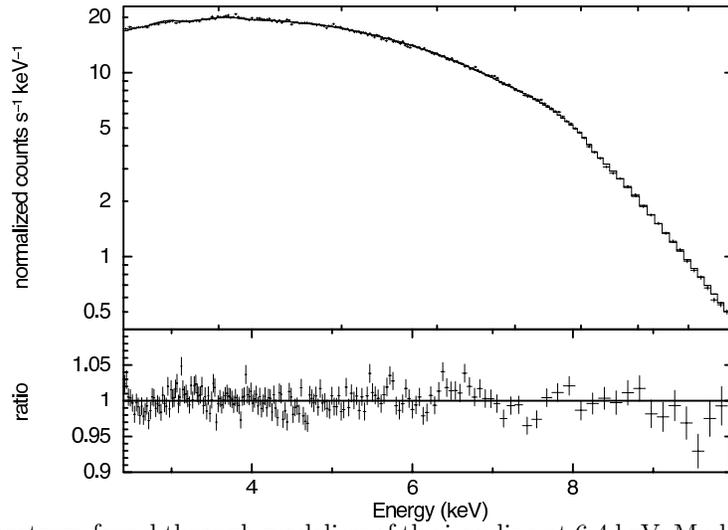}
\vspace{-1.0cm}
\end{center}
\caption{\footnotesize
{Best-fit spectrum found through modeling of the iron line at 6.4 keV.  Model: \texttt{ phabs*(powerlaw + kerrconv*reflionx)}. }}
\label{fig4}
\end{figure}

%-----------------------------------Figure End---------------------------------------------------------

\clearpage
%-----------------------------------Figure Start-------------------------------------------------------------
\begin{figure}[t]
\begin{center}
\includegraphics[scale=0.80]{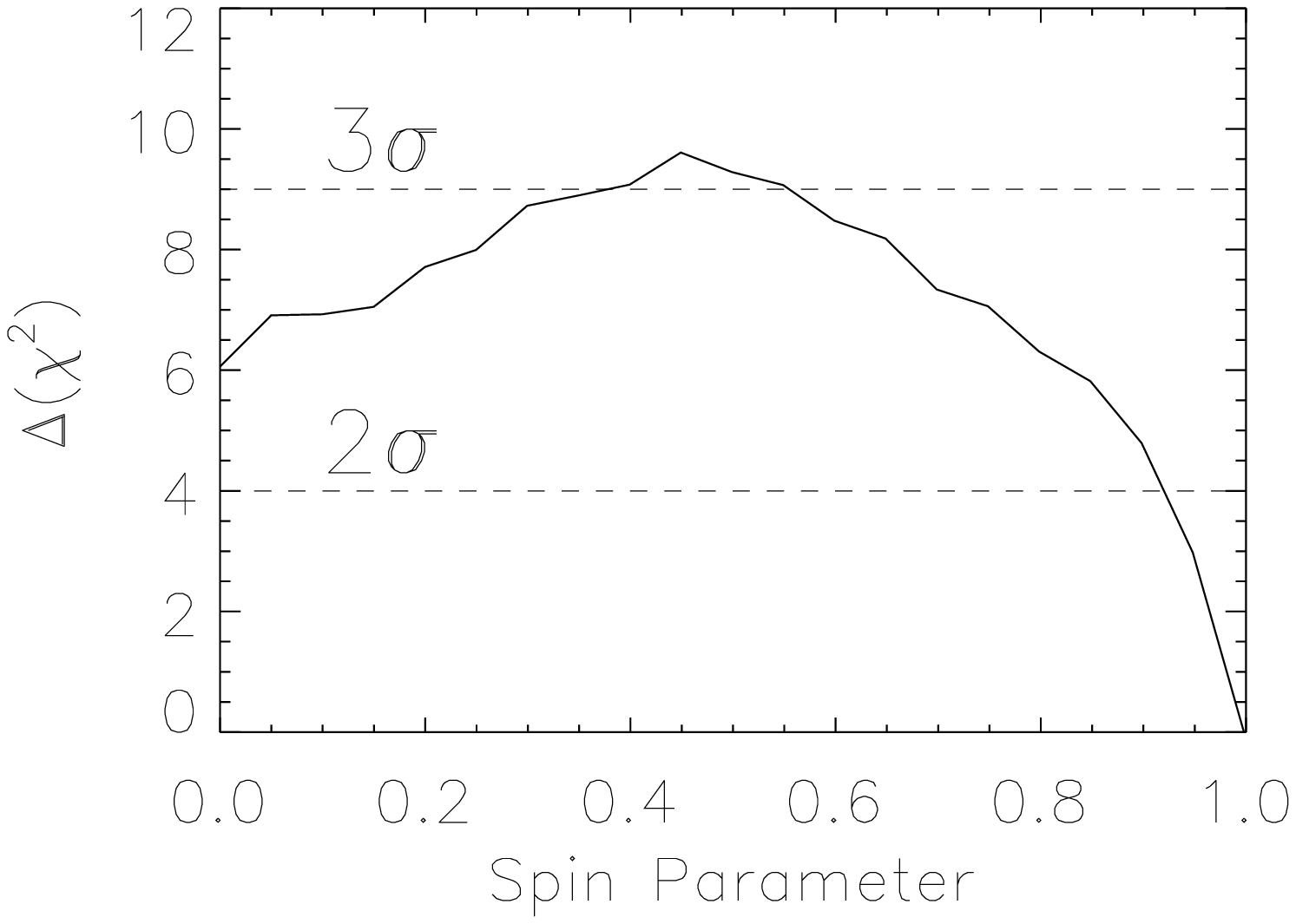}
\caption{\footnotesize
{The plot above shows the change in the goodness-of-fit statistic as a function of the black hole spin parameter, \emph{$\hat{a}$} for the model \texttt{ const.*phabs*(kerrdisk + kerrconv*pexriv)}.  Using the XSPEC \emph{steppar} command, 20 evenly-spaced values of \emph{$\hat{a}$} were frozen and all other parameters were allowed to float freely to find the best fit at that spin parameter.  The dotted lines indicate confidence intervals.}}

\label{fig5}
\end{center}
\end{figure}

%----------------------------------Figure End----------------------------------

\clearpage
%-----------------------------------Figure Start-------------------------------------------------------------
\begin{figure}[t]
\begin{center}
\includegraphics[scale=0.80]{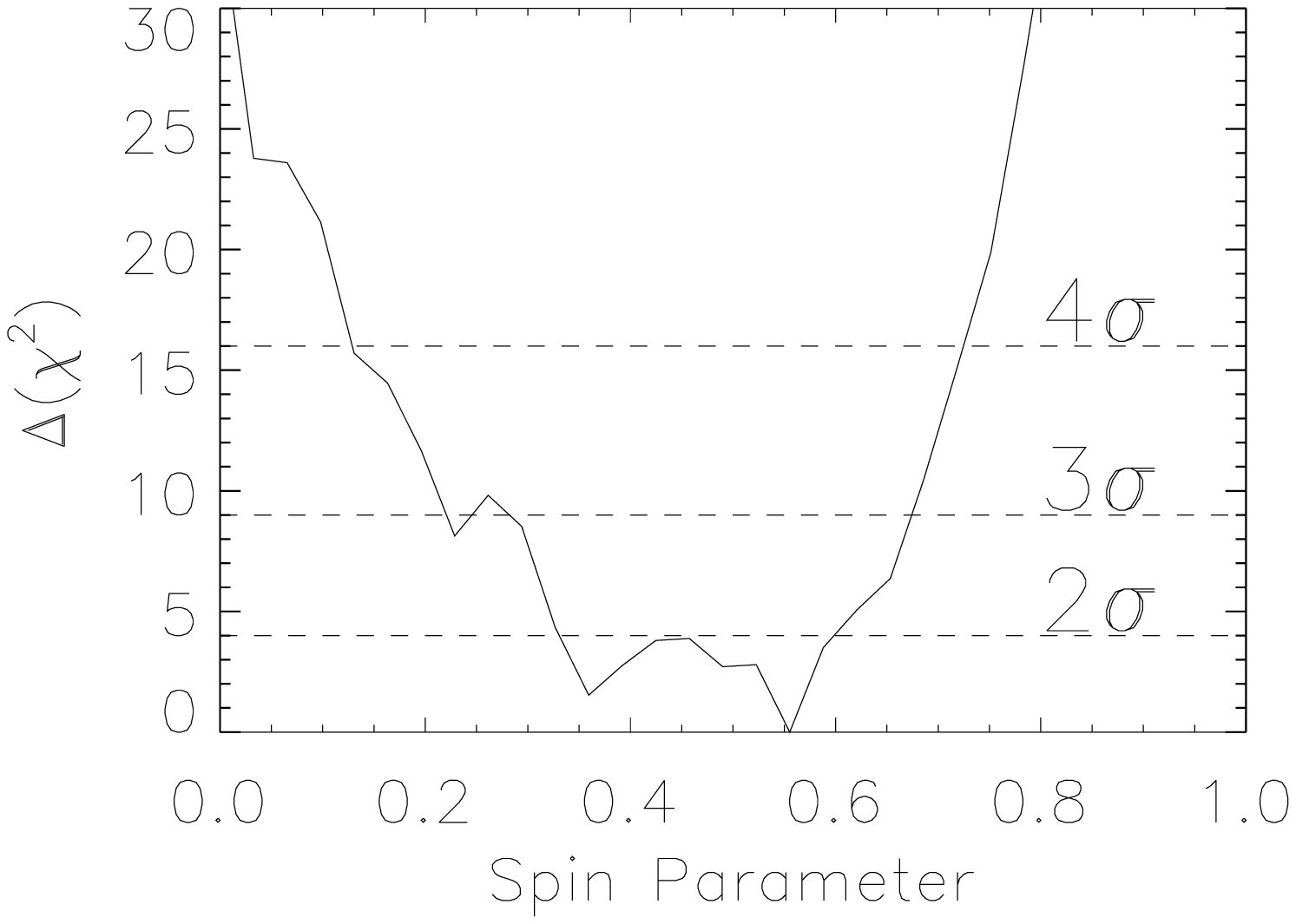}
\caption{\footnotesize
{The plot above shows the change in the goodness-of-fit statistic as a function of the black hole spin parameter, \emph{$\hat{a}$} for the model \texttt{phabs*(powerlaw+kerrconv*reflionx)}.  Using the XSPEC \emph{steppar} command, 30 evenly-spaced values of \emph{$\hat{a}$} were frozen and all other parameters were allowed to float freely to find the best fit at that spin parameter.  The dotted lines indicate confidence intervals.  Please note that only the XIS data was used in this model.}}

\label{fig6}
\end{center}
\end{figure}

%----------------------------------Figure End----------------------------------

%-----------------------------------Figure Start-------------------------------------------------------------
\begin{figure}[t]
\begin{center}
\includegraphics[scale=0.50]{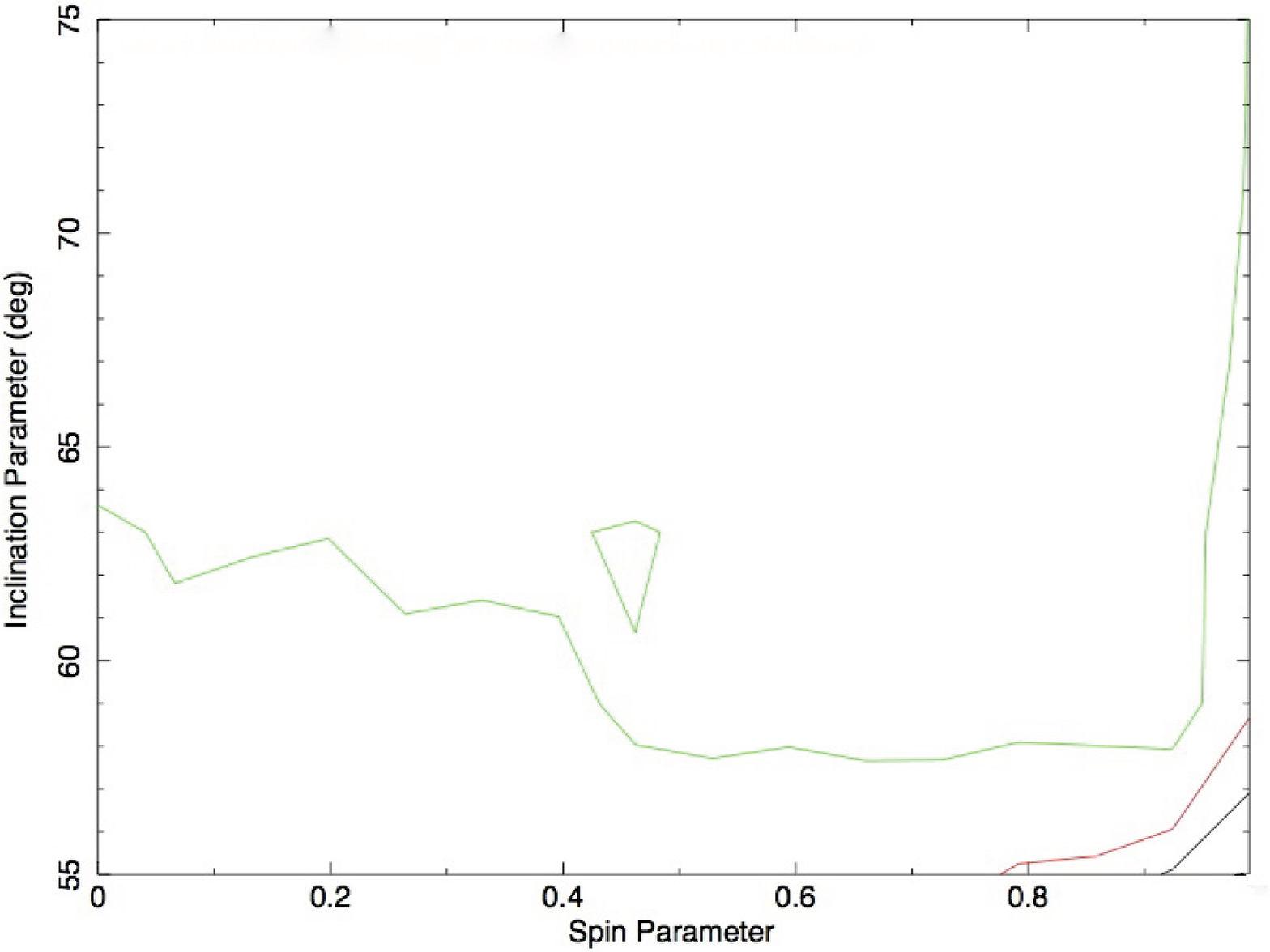}
\caption{\footnotesize
{Inclination versus spin contour plot for GRS 1915+105 using the \texttt{pexriv} model.  The 68, 90 and 99 per cent confidence range for two parameters of interest are shown in black, red and green, respectively.}}

\label{fig7}
\end{center}
\end{figure}

%----------------------------------Figure End----------------------------------

%-----------------------------------Figure Start-------------------------------------------------------------
\begin{figure}[t]
\begin{center}
\includegraphics[scale=0.50]{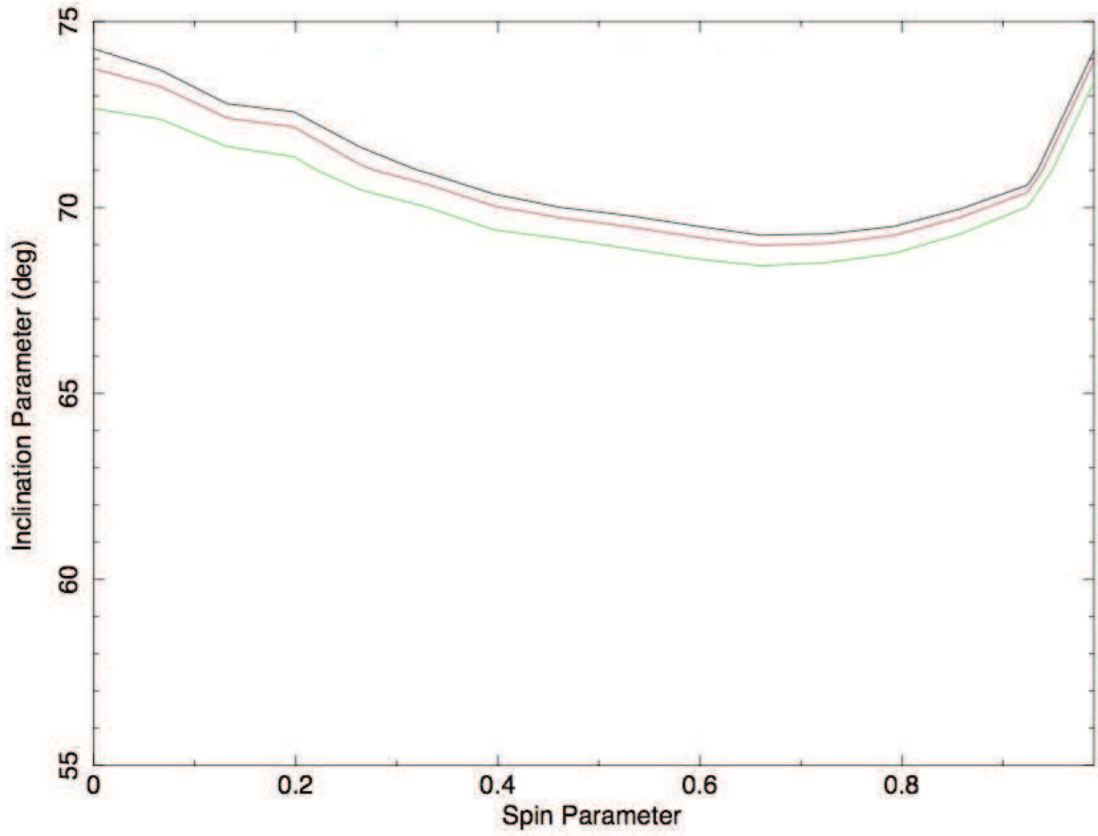}
\caption{\footnotesize
{Inclination versus spin contour plot for GRS 1915+105 using the \texttt{reflionx} model. The 68, 90 and 99 per cent confidence range for two parameters of interest are shown in black, red and green, respectively.}}

\label{fig8}
\end{center}
\end{figure}

%----------------------------------Figure End----------------------------------

\clearpage

\section*{References}
\noindent
Arnaud, K. A., 1996, ASP Conf. Series, 101, 17\\
Bardeen, J. M., Press, W. H., \& Teukolsky, S. A., 1972, ApJ, 178, 347\\
Beckwith K., \& Done C., 2004, MNRAS, 352, 352B\\
Belloni, T., et al., 2000, A\&A, 355, 271\\
Brenneman, L. W., \& Reynolds, C. S., 2006, ApJ, 652, 1028\\
\v{C}ade\v{z} A., \& Calvani, M., 2005, MNRAS, 363, 177\\
Castro-Tirado, A. J., Brandt, S., \& Lund, N. 1992, IAU Circ., 5590\\
Done, C., \& Gierli\'{n}ski , M., 2006, MNRAS, 367, 659\\ 
Dov\v{c}iak M., et al., 2004, ApJS, 153, 205D\\
Esin, A. A., McClintock, J. E., \& Narayan, R., 1997, ApJ, 489, 865\\
Fabian, A. C., et al., 1989, MNRAS, 238, 729\\
Fabian, A. C., Iwasawa, K., Reynolds, C. S., Young, A. J., 2000, PASP, 112, 1145\\
Fabian, A. C., \& Miniutti, G., 2005, in press, arXiv:astro-ph/0507409v1\\
Fender, R., et al. 1997, MNRAS, 290, L65\\
Fender, R., et al. 1999, MNRAS, 304, 865\\
Fender, R., \& Belloni, T., 2004, ARAA, 42, 317\\
Frank, J., King, A., \& Raine, D., 2002,  ``Accretion Power in Astrophysics'' (3rd. ed.)\\
George, I. M., \& Fabian, A. C., 1991 249, 352\\
Greiner, J., et al., 2001, A\&A, 373, L37\\
Kotani, T., et al., 2000, A\&A, 539, 413\\
Laor, A., 1991, ApJ, 376, 90\\
Lee, J. C., et al., 2001, X-ray Astronomy 2000: ASP Conference Series, 234, 231\\ 
Lee, J. C., et al., 2002, ApJ, 567, 1102\\
Magdziarz, P., \&  Zdziarski, A., 1995, MNRAS, 273, 837\\
Martocchia, A., et al., 2002, A\&A, 387, 215\\
Martocchia, A., et al., 2006, A\&A, 448, 677\\
Matt G., Fabian, A.C., Ross, R. R., 1996, MNRAS, 278, 1111\\
McClintock, J. E., et al., 2006, ApJ, 652, 518\\
McClintock, J. E., \& Remillard, R. A., 2009, arXiv:0902.3488v3\\
Middleton, M., et al., 2006, MNRAS, 373, 1004\\
Miller, J. M., 2007, ARAA, 45, 441\\
Miller, J. M., \& Homan, J., 2005, ApJ, 618, L107\\
Miller, J. M., et al., 2006a, ApJ, 653, 525\\
Miller, J. M., et al., 2006b, ApJ, 652, L113\\
Miller, J. M., et al., 2008a, ApJ, 680, 1359\\
Miller, J. M., et al., 2008b, ApJ, 679, L113\\
Miller, J. M., Reynolds, C. S., Fabian, A. C., Miniutti, G., \& Gallo,
L. C., 2009, ApJ, 697, 900\\
Nielsen, J., \& Lee, J. C., 2009, Nature, in press\\
Nowak, M. A., 1995, PASP, 107, 1207\\
Park, S. Q., et al., 2004, ApJ, 610, 378\\
Reis, R.C., et al., 2008, MNRAS, 387, 1489\\
Reis, R. C., et al., 2009, MNRAS, in press (arXiv:0902.1745)\\
Reynolds, C. S.,  \& Begelman, M. C., 1997, ApJ, 488, 109R\\
Reynolds, C. S., \& Fabian, A. C., 2008, ApJ, 675, 1048\\
Reynolds, C. C., \& Nowak, M. A., 2003, PhR, 377, 389\\
Rykoff, E., Miller, J. M., Steeghs, D., \& Torres, M. A. P., 2007, ApJ, 666, 1129\\
Ross, R.R., Fabian, A.C., 1999, MNRAS, 306, 461\\
Ross, R. R., \& Fabian, A. C., 2005, MNRAS, 358, 211\\
Shafee, R., et al., 2006, ApJ, 636, L113\\
Shafee, R., et al., 2008, ApJ, 687, L25\\
Sobczak, G. J., et al., 2000, ApJ, 544, 993\\
Thorne, K. S., 1974, ApJ, 191, 507T\\
Tomsick, J. A., et al., 2008, ApJ, 680, 593\\
Ueda, Y., Yamaoka, K., \& Remillard, R., 2009, ApJ, 695, 888\\

\end{document}